\documentclass[modern,times]{aastex62}%

\usepackage{amsmath}          

\usepackage{graphicx}         
\usepackage{dcolumn}          
\usepackage{xspace}           
\usepackage{xifthen}          
\usepackage{mathrsfs}         
\usepackage{txfonts}          
\usepackage{textcomp}
\usepackage{nicefrac}
\usepackage[version=4]{mhchem}

\providecommand*{\wn}{cm$^{-1}$\xspace}

\newcommand{\mrm}[1]{\ensuremath{\mathrm{#1}}}
\newcommand{\mcl}[3]{\multicolumn{#1}{#2}{#3}}

\newcommand{\bra}[2][]{%
  \ifthenelse{\isempty{#1}}%
  {\ensuremath{\left\langle #2 \right\rvert}}%
  {\ensuremath{_{#1}\left\langle #2 \right\rvert}}}
\newcommand{\ket}[2][]{%
  \ifthenelse{\isempty{#1}}%
  {\ensuremath{\left\lvert #2 \right\rangle}}%
  {\ensuremath{\left\lvert #2 \right\rangle_{#1}}}}

\makeatletter
\def\@dotsep{4.5}
\makeatother

\newcolumntype{.}{D{.}{.}{-1}}
\newcolumntype{d}[1]{D{.}{.}{#1}}
\renewcommand{\thefootnote}{\fnsymbol{footnote}}

\received{\today}
\revised{revision date}
\accepted{acceptance date}
\published{published date}
\submitjournal{\apjs}

\shorttitle{THz spectroscopy of \ce{NHD}}
\shortauthors{Bizzocchi et al.}

\begin{document}

\title{Submillimetre and far-infrared spectroscopy of monodeuterated amidogen radical 
       (\ce{NHD}): improved rest-frequencies for astrophysical observations}

\author{Luca Bizzocchi}
\affiliation{Center for Astrochemical Studies, 
  Max-Planck-Institut f\"ur extraterrestrische Physik, \\
  Gie\ss enbachstr.~1, 85748 Garching bei M\"unchen (Germany)}

\author{Mattia Melosso}
\affiliation{Dipartimento di Chimica ``G.~Ciamician'',
  Universit\`a di Bologna, \\
  via F.~Selmi 2, 40126 Bologna (Italy)}

\author{Barbara Michela Giuliano}
\affiliation{Center for Astrochemical Studies, 
  Max-Planck-Institut f\"ur extraterrestrische Physik, \\
  Gie\ss enbachstr.~1, 85748 Garching bei M\"unchen (Germany)}  

\author{Luca Dore}
\affiliation{Dipartimento di Chimica ``G.~Ciamician'',
  Universit\`a di Bologna, \\
  via F.~Selmi 2, 40126 Bologna (Italy)}

\author{Filippo Tamassia}
\affiliation{Dipartimento di Chimica Industriale ``Toso Montanari'', 
  Universit\`a di Bologna, \\
  viale del Risorgimento~4, 40136 Bologna (Italy)}

\author{Marie-Aline Martin-Drumel}
\affiliation{Institut des Sciences Mol\'eculaires d'Orsay (ISMO), 
  CNRS, Univ. Paris-Sud, Universit\'e Paris-Saclay, F-91405 Orsay, France}

\author{Olivier Pirali}
\affiliation{Institut des Sciences Mol\'eculaires d'Orsay (ISMO), 
  CNRS, Univ. Paris-Sud, Universit\'e Paris-Saclay, F-91405 Orsay, France}
\affiliation{SOLEIL Synchrotron, AILES beamline, l'Orme des Merisiers, Saint-Aubin, 
  F-91190 Gif-sur-Yvette, France}

\author{Laurent Margul\`es}
\affiliation{Universit\'e Lille, CNRS, UMR8523 - PhLAM - Physique des Lasers Atomes et Mol\'ecules, 
  F-59000 Lille, France}

\author{Paola Caselli}
\affiliation{Center for Astrochemical Studies, 
  Max-Planck-Institut f\"ur extraterrestrische Physik, \\
  Gie\ss enbachstr.~1, 85748 Garching bei M\"unchen (Germany)}

\correspondingauthor{Luca Bizzocchi}
\email{bizzocchi@mpe.mpg.de}

\correspondingauthor{Mattia Melosso}
\email{mattia.melosso2@unibo.it}


\begin{abstract}
\noindent
Observations of ammonia in interstellar environments have revealed high levels of deuteration, 
and all its D-containing variants, including \ce{ND3}, have been detected in 
cold prestellar cores and around young protostars. 
The observation of these deuterated isotopologues is very useful to elucidate the chemical 
and physical processes taking place during the very early stages of star formation, as the 
abundance of deuterated molecules is highly enhanced in dense and cold gas.
Nitrogen hydride radicals are key species lying at the very beginning of the reaction pathway 
leading to the formation of \ce{NH3} and organic molecules of pre-biotic interest, but 
relatively little information is known about their D-bearing isotopologues.
To date, only \ce{ND} has been detected in the interstellar gas.
To aid the identification of further deuterated nitrogen radicals, we have thoroughly 
re-investigated the rotational spectrum of \ce{NHD} employing two different instruments: 
a frequency-modulation submillimetre spectrometer operating in the THz region and a 
synchrotron-based Fourier Transform infrared spectrometer operating in the 50--240 \wn 
frequency range. 
\ce{NHD} was produced in a plasma of \ce{NH3} and \ce{D2}. 
A wide range of rotational energy levels has been probed thanks to the observation of 
high $N$ (up to~15) and high $K_a$ (up to~9) transitions. 
A global analysis including our new data and those already available in the literature 
has provided a comprehensive set of very accurate spectroscopic parameters. 
A highly reliable line catalogue has been generated to assist archival data searches and 
future astronomical observations of \ce{NHD} at submillimetre and THz regimes.
\end{abstract}



\section{Introduction} \label{sec:intro}
\indent\indent
Nitrogen is amongst the most abundant elements in the Universe and is a constituent of 
essentially all molecules which are important for life as we know it.
Nitrogen-bearing species are also ubiquitous in the interstellar medium (ISM) and 
circumstellar envelopes where about 60 such molecules have been detected to date 
(\citealt{McGuire-ApJS18-Census}, see also \texttt{http://astrochymist.org/} and 
CDMS\footnote{%
  Cologne Database for Molecular Spectroscopy, \texttt{https://cdms.astro.uni-koeln.de/classic/}},
\citealt{Endres-JMS16-CDMS}).
The main molecular reservoir of nitrogen in the ISM is believed to be the homonuclear form
\ce{N2}.
Observation of this molecule, however, is extremely challenging: first, the lack of a permanent 
dipole moment due its centrosymmetric nature prevents this species from having a rotational
signature; second, its vibrational mode is infrared inactive.
Consequently, a single interstellar detection of \ce{N2} through its
electronic spectrum in the far ultraviolet has been reported to date \citep{Knauth-Nat04-N2}.
The study of nitrogen chemistry in space thus mainly relies on observations of other
molecules.

Light nitrogen hydrides as imidogen (\ce{NH}), amidogen (\ce{NH2}), and ammonia (\ce{NH3}) 
are involved in the very first steps of the reaction network which leads to the formation 
of complex species, and are thus of primary importance to constrain the models describing 
nitrogen chemistry in the ISM.
They are formed by a chain of hydrogen abstraction reactions starting from:
\begin{equation} \label{eq:Nchain}
\text{N}^+ + \text{H}_2 \longrightarrow \text{NH}^+ + \text{H} \,,
\end{equation}
and that produce the ions \ce{NH2+}, \ce{NH3+}, and \ce{NH4+}, which are connected to the 
corresponding neutral hydrides by dissociative recombination reactions. 
\citep{LeGal-AA14-Nhyd,Dislaire-AA12-NH}.

The ion--neutral process \eqref{eq:Nchain} is endothermic and is thus very 
sensitive to the \ce{H2} ortho-to-para ratio in the cold ($T < 20$\,K) and dense ISM.
Also, if \ce{H2} is replaced with its singly deuterated form \ce{HD}, the endothermicity 
of \eqref{eq:Nchain} is reduced (see, e.g. \citealt{Cerni-ApJL13-NH3D+}) thus favouring,
in principle, the formation of ND$^+$ and further D-susbtituted species 
(the so called ``fractionation'').
However, given that the \ce{HD}/\ce{H2} ratio keeps about constants to $\sim 10^{-5}$
\citep{Linsky-SSR07-D_H}, this route is rather inefficient, and it is likely that 
reactions of neutral species with \ce{H2D+}, which abund in the dense and cold environment,
could be efficient drivers of D-fractionation \citep[e.g.][]{Sipila-AA19-Dchem}.
This process is triggered by the favorable thermochemistry (deuterated isotopologues 
typically have lower zero-point energy than the corresponding parent species), and if
suitable chemical conditions are met (i.e. freeze-out of CO onto dust grains 
\citealt{Caselli-ApJ99-COdep,Caselli-ApJ02-L1544i}), it can be very effective in enhancing 
the abundances of D-bearing species in cold and dense environments.

Albeit statistically improbable owing to the relatively low cosmic elemental abundance of 
deuterium ($\mrm{D}/\mrm{H}\sim 1.6\times 10^{-5}$), multiply deuterated variants of 
ammonia are known to be present in space (see, e.g., \citealt{Roueff-AA05-NH3}), 
including the triply substituted \ce{ND3} \citep{Lis-ApJ02-ND3,vdTak-AA02-ND3} whose 
estimated abundance ($\mrm{ND}_3/\mrm{NH}_3\sim 10^{-3}-10^{-4}$) implies an isotopic 
enhancement of up to 12~orders of magnitude \citep{Cecc-P&P07-Dchem}.

Interestingly enough, among the ammonia progenitors, only ND has been detected in space so 
far \citep{Bacmann-AA10-L1544, Bacmann-AA16-L16293E} even though  significant abundances 
of \ce{NHD} and \ce{ND2} are predicted by chemical models \citep{Roueff-AA05-NH3}.
To explain such lack of detection, one has to point out that nitrogen hydride radicals are 
elusive species.
Being light molecules, their rotational spectral features are located in the submillimetre
(submm) to far-infrared (FIR) ranges, where Earth-based observations face the limitation 
of the high atmospheric opacity.
This spectral window has been widely opened in the last decade by the 
\emph{Herschel Space Telescope} campaign \citep{Pilb-AA10-HSO}, thus triggering an 
increased interest for the spectroscopic properties of light molecules, 
hydrides in particular.
Nowadays, with ALMA at its full capabilities, wide portions of the submm regime up to 
$\sim$1\,THz have become more accessible from the ground, and further perspectives can be 
foreseen with the improvement of the SOFIA airborne observatory
\citep[see, e.g.][]{Yorke-SPIE18-SOFIA},
in particular, with the upcoming commissioning of the 4GREAT receiver, which is designed to 
replicate most of the spectral coverage of the \emph{Herschel}/HIFI instrument.
Still, high-frequency observations remain a challenging task and a successful 
detection of the spectral features of light molecular tracers critically depends on the 
accuracy of the corresponding rest-frequencies. 

From a spectroscopic point of view, the parent species \ce{NH2} has been widely studied in 
the laboratory over the past 40~years 
\citep{Davies-CPL76-NH2,Hills-JMS82-NH2,Charo-ApJL81-NH2,Tonoo-JCP97-NH2,Muller-JMS99-NH2,%
       Gendr-JMSt01-NH2,Martin-JPCA14-NH2}, 
while much less extensive data sets are available for the other amidogen isotopologues.
Driven by the perspectives of the current observational facilities, the symmetric species 
$^{15}$\ce{NH2} and \ce{ND2} species have been recently re-investigated at high resolution 
in the millimetre (mm) and FIR domains \citep{Margules-AA16-15NH2,Melosso-ApJS17-ND2}, 
and the rotational spectrum of the fully substituted $^{15}$\ce{ND2} variant has been 
reported for the first time \citep{Melosso-JQSRT19-15ND2}.
A thorough spectroscopic study is still missing for the asymmetric singly-substituted \ce{NHD}
species, which is likely to be the most abundant deuterated variant of amidogen in the ISM.
In the early '80s, the hyperfine structure of a few lines of \ce{NHD} were observed by 
Microwave-Optical Double Resonance (MODR) technique \citep{Brown-ApJ80-NHD,Steimle-JCP80-NHD}, 
whereas the first extensive investigation of its rotational spectrum was accomplished much 
later by \citet{Morino-JMA97-NH2} who employes a Fourier-Transform (FT) FIR spectrometer.
Almost simultaneously, \citet{Kobay-JCP97-NH2} published an analogous study in the mm/submm
domain using a microwave absorption spectrometer.
In the former case, the spectral resolution did not allow the observation of the hyperfine
structure but the spectral coverage (103--363\,\wn) led to a detailed centrifugal distortion
analysis.
In the latter case, hyperfine splittings were resolved for many transitions, and several
spin-spin and spin-rotation constants were determined for $^{14}$N, H, and D.

In this work, we report on: ($i$) the extension towards the THz regime of the rotational 
spectrum of \ce{NHD} with high measurement accuracy, i.e., better than 100\,kHz, and 
($ii$) a re-investigation of its FIR spectrum, recorded at higher resolution (0.001\,\wn) 
using synchrotron radiation at the AILES beamline of the SOLEIL synchrotron.
Newly observed and previously recorded data were analysed together in order to produce a
unique set of highly accurate spectroscopic constants.
The structure of the paper is the following: In \S~\ref{sec:exp}, we describe the laboratory
measurements; in \S~\ref{sec:anal}, we provide an account of the data used in the analysis 
and give a brief description of the Hamiltonian employed to compute the rotational, 
fine-structure, and hyperfine energies.
We discuss the results in \S~\ref{sec:res}, and draw our conclusions in \S~\ref{sec:conc}.

\section{Experiments} \label{sec:exp}
\indent\indent
High-resolution pure rotational spectral data of \ce{NHD} have been collected in the submm 
region at the Centre for Astrochemical Studies of the Max-Planck-Institut f\"ur 
extraterrestrische Physik in Garching (CAS@MPE) and in the FIR domain at the AILES beamline 
of the SOLEIL synchrotron in Gif-sur-Yvette.

The measurements in Garching have been carried out using the CASAC (CAS Absorption Cell)  
spectrometer associated to a discharge absorption cell for the studies of reactive species 
\citep{Bizz-ApJL18-15NH}.
A detailed description of the instrument has been given earlier \citep{Bizz-AA17-HOCO+}; 
we report here only a few key details which apply to the present investigation.
Several solid-state multiplier/amplifier chains (Virginia Diodes), driven by a radio-frequency 
synthesizer, have been used as radiation sources.
Below 1.1\,THz we have employed the WR9.0SGX module, working in the 80--125\,GHz interval, 
associated to a series of active and passive harmonic doublers/triplers in cascade, 
thus achieving a frequency multiplication factors as high as~9.
Measurements above this frequency have been performed with a stand-alone active multiplier 
system operating in the 1.1--1.2\,THz (AMC-680) interval.
Accurate frequency and phase stabilisation is achieved by locking the parent centimetre 
synthesizer to a Rb atomic clock.
A closed-cycle He-cooled InSb hot electron bolometer operating at 4\,K (QMC) is used as 
a detector.
The measurements have been performed using the frequency modulation (FM) technique 
to improve the signal-to-noise ratio (S/N).
In each frequency range, the FM depth was chosen to approximately match the Doppler 
half-width of the absorption signals, i.e, 500\,kHz at 0.5\,THz, and 1300\,kHz at 1.2\,THz.
The carrier signal is modulated by a 33.3\,kHz sine-wave and the detector output 
is demodulated at twice this frequency using a phase sensitive amplifier ($2f$ detection).
In this way, the second derivative of the actual absorption profile is recorded by the 
computer controlled acquisition system.
An additional S/N improvement is achieved by filtering the signal into a $RC$ circuit.

The \ce{NHD} radical was produced in a glow discharge (70\,mA, $\sim$ 1.1\,kV) of a 1:2 mixture
of \ce{NH3} (4\,\textmu bar) and D$_2$ (8\,\textmu bar) diluted in Ar buffer (total pressure 
$\sim 25$\,\textmu bar), with the cell kept at $\sim 180$\,K by cold vapour/liquid \ce{N2} 
circulation.
Besides preventing cell overheating, the cooling was found to be critical to improve 
the amidogen production in the plasma.
However, below the indicated temperature, \ce{NHD} signals decreased due to massive ammonia 
condensation on the cold cell walls.
The spectrum was recorded in selected frequency regions from 430\,GHz to 1.2\,THz\@.
Slow (0.67--1.67\,MHz\,s$^{-1}$) back-and-forth scans around target lines were performed 
employing a frequency step of 10--25\,kHz and a time constant $RC$ = 3\,ms\@.

The FIR spectrum of \ce{NHD} has been recorded at the synchrotron facility SOLEIL using a 
Bruker IFS125HR FTIR spectrometer exploiting the bright synchrotron radiation extracted by 
the AILES beamline. 
The spectrum results from the co-addition of 104~scans recorded at the ultimate resolution 
of the instrument, $R = 0.001$\,\wn in terms of Bruker definition, in the 50--250\,\wn\ range. 
A 6\,\textmu m Mylar beamsplitter and a 4.2\,K liquid helium cooled bolometer were used.
The \ce{NHD} radical was produced using a post-discharge set-up successfully employed in 
the past to produce the \ce{NH2} and NH radicals as well as their $^{15}$N isotopic variants 
\citep{Martin-JPCA14-NH2,Margules-AA16-15NH2,Baill-AA12-15NH}. 
In this configuration, a radiofrequency discharge cell is connected perpendicularly to the 
center of a White-type multipass absorption cell. 
In the present work, a 2.5\,m baselength White-type cell allowing 150\,m of absorption 
path length was used \citep{Pirali-JCP12-And}.
The absorption cell was separated from the interferometer by two polypropylene films of 
60\,\textmu m thickness. 
The spectrometer was pumped down to $10^{-4}$\,mbar by means of a turbomolecular pump, while 
the flow in the post-discharge experiment was ensured by a booster pump.
The \ce{NHD} radical was produced by a 1000\,W radiofrequency discharge \citep{Martin-RSI11-AILES} 
in a \mbox{\ce{ND3} + \ce{H2}} mixture at partial pressures of 30 and 15\,\textmu bar, respectively. 
Production of \ce{NHD}, as well as \ce{ND2}, were found to be more efficient when the gas mixture 
was injected at one end of the absorption cell while pumping connections were set at 
the other end of that cell, so that no gas was directly injected through the RF discharge cell. 
Such configuration enabled a propagation of the plasma into the absorption cell. 
A discharge-off spectrum was also recorded to quickly identify the lines arising from 
transient species.

\section{Observed spectra and analysis} \label{sec:anal}
\indent\indent
Monodeuterated amidogen is an asymmetric-top rotor belonging to the $C_s$ point group symmetry.
Because of the presence of one unpaired electron in a $a^{\prime\prime}$ non-bonding orbital,
\ce{NHD} is a radical with a $\tilde{X}\,^2A^{\prime\prime}$ electronic ground state.
The electric dipole moment lies in the $ab$ principal symmetry plane with components 
$\mu_a = 0.67$\,D
and $\mu_b = 1.69$\,D \citep{Brown-MP79-NH2}.
The rotational spectrum of \ce{NHD} is very complex due to the coupling of the molecular rotation
with various electronic and nuclear spin angular momenta.
The electron spin ($S=\nicefrac{1}{2}$) couples with the rotational angular momentum and splits 
each rotational level with $N > 0$ into two fine-structure sublevels having 
$J = N + \nicefrac{1}{2}$ and $J = N - \nicefrac{1}{2}$ quantum numbers.
Each of these sublevels is further split by hyperfine interactions due to the nuclear spins
of $^{14}$N ($I=1$), H ($I=\nicefrac{1}{2}$), and D ($I=1$).
The chosen angular momentum coupling scheme
can be summarised as:
\begin{subequations} \label{eq:coup}
  \begin{eqnarray} \label{eq:subcoup}
  \mathbf{J}    &=&  \mathbf{N}   + \mathbf{S}           \, \\
  \mathbf{F}_1  &=&  \mathbf{J}   + \mathbf{I}_\mrm{N}   \, \\
  \mathbf{F}_2  &=&  \mathbf{F}_1 + \mathbf{I}_\mrm{H}   \, \\
  \mathbf{F}    &=&  \mathbf{F}_2 + \mathbf{I}_\mrm{D}   \,
  \end{eqnarray}
\end{subequations}
where $\mathbf{N}$ and $\mathbf{S}$ are the rotational and electron spin angular momenta, 
whereas $\mathbf{I}_X$ ($X = \mrm{N},\mrm{D},\mrm{H}$) represents the various nuclear spin
angular momenta.
In Eqs.~\eqref{eq:coup}, the nuclear spins are ordered according to the approximate magnitude 
of their hyperfine coupling.

\begin{figure*}
  \includegraphics[angle=0,width=0.95\textwidth]{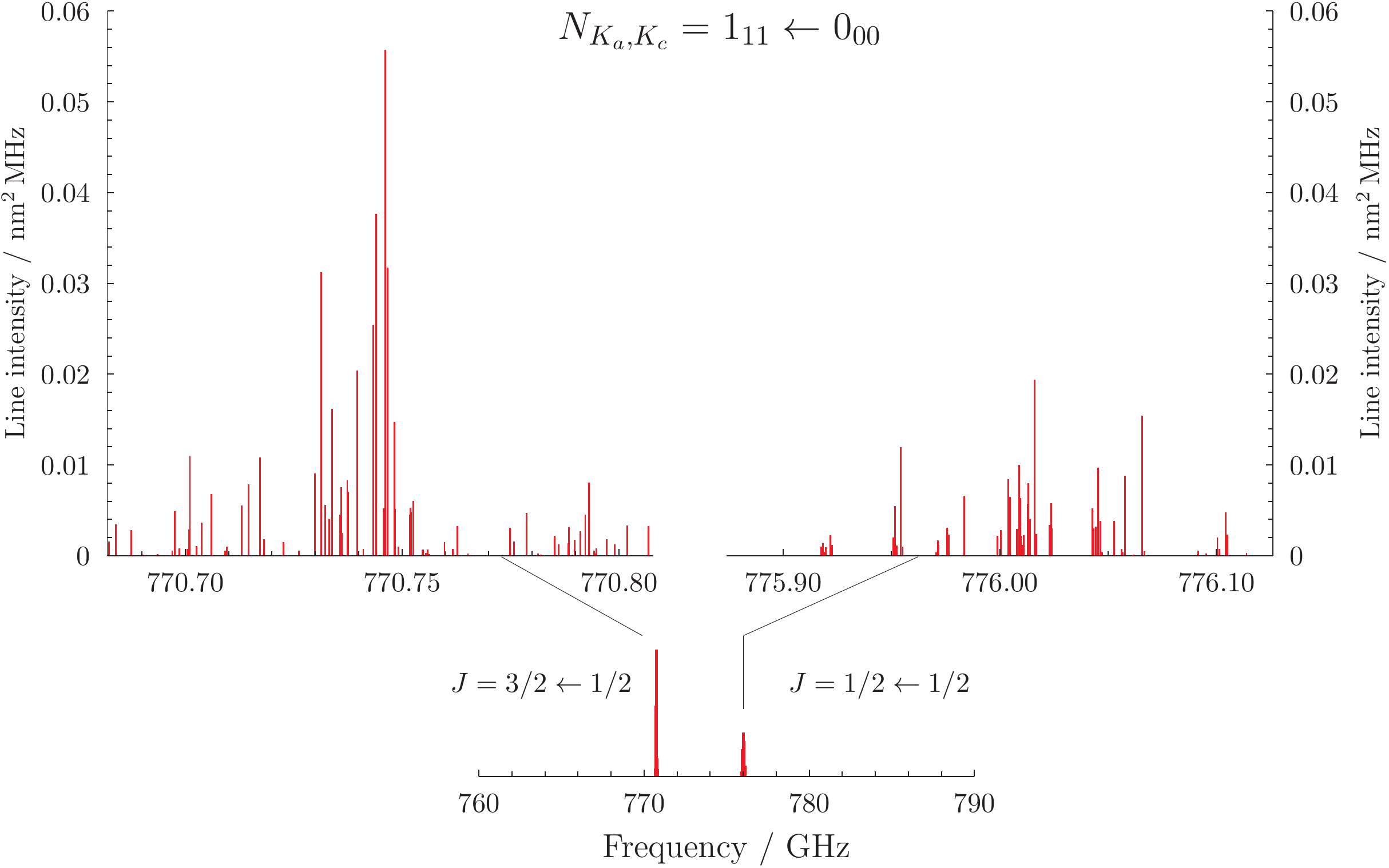}
  \caption{Stick spectrum of the $N_{Ka,Kc} = 1_{11}-0_{00}$ rotational transition of \ce{NHD} simulated at 298\,K. \label{fig:111-000}}
\end{figure*}

The effective Hamiltonian can be expressed as a sum of several energy contributions as in
the following \citep{Morino-JMA97-NH2,Kobay-JCP97-NH2}:
\begin{equation} \label{eq:Heff}
\tilde{H}_\mrm{eff} = \tilde{H}_\mrm{rot} + \tilde{H}_\mrm{sr} + \tilde{H}_\mrm{hfs} \,.
\end{equation}
Here, $\tilde{H}_\mrm{rot}$ is the Watson $A$-reduced Hamiltonian in its $I^r$ representation 
\citep{Watson-in77-H}, which includes the rotational energy and the centrifugal distortion
(up to the $J^{12}$ terms in the present analysis).
The fine-structure Hamiltonian $\tilde{H}_\mrm{sr}$ contains the electron spin-rotation terms 
($\boldsymbol{\epsilon}$ tensor) and their centrifugal dependencies.
The $\tilde{H}_\mrm{hfs}$ term is the hyperfine-structure Hamiltonian which includes electronic
spin-nuclear spin interactions ($a_F$, $\mathbf{T}$) and nuclear spin-nuclear spin 
coupling ($\mathbf{C}$) for all the three nuclei, and the electric quadrupole coupling 
contributions ($\boldsymbol{\chi}$) for the $^{14}\mrm{N}$ and $\mrm{D}$ nuclei, 
having $I\geq 1$. 
The spectral computation was performed using the SPFIT/SPCAT suite of programs
\citep{Pick-JMS91-calpgm} which implements all the matrix elements of the 
Hamiltonian~\eqref{eq:Heff} in the coupling scheme~\eqref{eq:coup}.

Unlike the symmetric species \ce{NH2} and \ce{ND2}, for which a $\pi$ rotation around the 
$b$ principal axis exchanges two identical particles, there are no spin statistics in the 
case of \ce{NHD}\@.
Consequently, all fine and hyperfine sublevels, up to a maximum of~36, are allowed for each 
$N_{K_a,K_c}$ rotational level.
The electric dipole selection rules \citep{Gordy-1984} on this sublevels manifold produce 
very complex hyperfine patterns for each rotational transition due to the additional 
fine/hyperfine selection rules $\Delta J=0,\pm 1$, $\Delta F_1=0,\pm 1$, $\Delta F_2=0,\pm 1$, 
and $\Delta F=0,\pm 1$. 
The most intense components have $\Delta J=\Delta F_1=\Delta F_2=\Delta F=\Delta N$\@. 
An example of the complexity is given in Figure~\ref{fig:111-000}, which illustrates the 
above described interactions for the fundamental $b$-type transition $N_{K_a,K_c} = 1_{11}-0_{00}$.
Each line of the fine-structure doublet, separated by some GHz, is then split in a multitude
of hyperfine components, roughly gathered in loose triplets, due to the dominant effect 
of electron spin--nuclear spin coupling due to the $^{14}$N nucleus.

\begin{table}[ht!]
  \caption{List of \ce{NHD} rotational transitions with HFS recorded in the submillimetre
    region.\label{tab:submm-lin}}
  \vspace{5mm}
  \begin{center}
    \begin{tabular}{cl d{3} cc}
      \hline\hline  \\[-2ex]
      \mcl{1}{c}{line}                            &
      \mcl{1}{c}{type$^a$}                        &
      \mcl{1}{c}{unsplit freq. (MHz)} &
      \mcl{1}{c}{no. of FS lines}                 &
      \mcl{1}{c}{no. of HFS lines}                \\[1ex]
      \hline  \\[-2ex]
      $1_{1,0} - 1_{0,1}$  &  $^bQ_{+1,-1}$  &   432580.3   &  4  &  44  \\[0.5ex]
      $2_{0,2} - 1_{1,1}$  &  $^bR_{-1,+1}$  &   456235.5   &  2  &   5  \\[0.5ex]
      $2_{1,1} - 2_{0,2}$  &  $^bQ_{+1,-1}$  &   515448.6   &  2  &  19  \\[0.5ex]
      $3_{1,2} - 3_{0,3}$  &  $^bQ_{+1,-1}$  &   656654.9   &  2  &  30  \\[0.5ex]
      $1_{1,1} - 0_{0,0}$  &  $^bR_{+1,+1}$  &   772506.7   &  2  &  15  \\[0.5ex]
      $4_{1,3} - 4_{0,4}$  &  $^bQ_{+1,-1}$  &   869336.3   &  2  &   7  \\[0.5ex]
      $3_{0,3} - 2_{1,2}$  &  $^bR_{-1,+1}$  &   902924.7   &  2  &  12  \\[0.5ex]
      $3_{2,1} - 3_{1,2}$  &  $^bQ_{+1,-1}$  &  1020828.9   &  2  &   5  \\[0.5ex]
      $6_{2,4} - 6_{1,5}$  &  $^bQ_{+1,-1}$  &  1062344.6   &  2  &   6  \\[0.5ex]
      $2_{1,2} - 1_{0,1}$  &  $^bR_{+1,+1}$  &  1112159.2   &  2  &   4  \\[0.5ex]
      $3_{1,3} - 2_{1,2}$  &  $^aR_{0,+1}$   &  1122794.6   &  1  &   1  \\[0.5ex]
      $5_{1,4} - 5_{0,5}$  &  $^bQ_{+1,-1}$  &  1157071.0   &  1  &   6  \\[0.5ex]
      $3_{0,3} - 2_{0,2}$  &  $^aR_{0,+1}$   &  1199330.9   &  2  &   6  \\[0.5ex]
      \hline\hline \\[-2ex]
    \end{tabular}
  \end{center}
  $^a$ The symbol $^xM_{\delta K_a,\delta K_c}$ is used to label in a compact form the transition
  type for an asymmetric rotor: $x$ indicates the dipole moment component involved, 
  $M = P,Q,R$ is the symbol for the transitions with $\Delta N = -1,0,+1$, respectively, 
  and $\delta K_a$ and $\delta K_c$ refer to the (signed) change in the $K_a$ and $K_c$ 
  pseudo-angular quantum numbers \citep{Gordy-1984}.
\end{table}

\subsection{Submillimetre spectrum} \label{sec:submm}
\indent\indent
The spectral recordings performed with the CASAC spectrometer covered the frequency 
interval from 430\,GHz to 1200\,GHz, exploring the region above 520\,GHz for the first time.
Since \ce{NHD} is a very light molecule, the rotational transitions are quite isolated in the 
spectrum and separated by several tens of GHz.
The transitions recorded in this frequency domain are summarized in  Table~\ref{tab:submm-lin},
which also reports the hypothetically unsplit frequency of the rotational lines and the 
number of assigned hyperfine components.

\begin{figure*}
  \includegraphics[angle=0,width=0.95\textwidth]{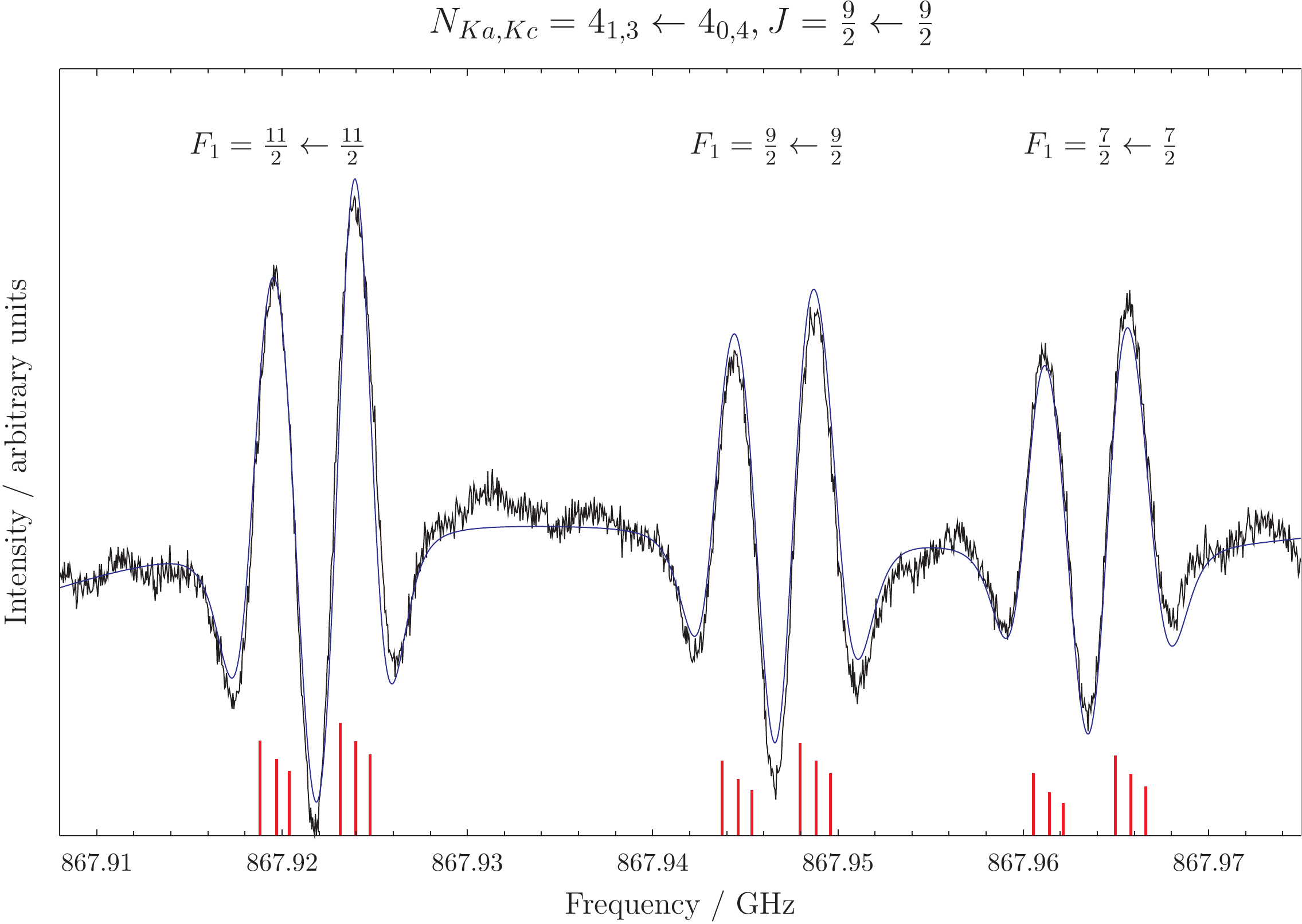}
  \caption{Recording (black trace) of the 
    $N_{K_a,K_c} = 4_{1,3}\leftarrow 4_{0,4}, J = \nicefrac{9}{2}\leftarrow  \nicefrac{9}{2}$
    fine-structure transitions showing six resolvable components.
    Integration time: 390\,s; $RC = 3$\,ms; scanning rate: 1.3\,MHz/s; modulation
    depth: 850\,kHz.
    The blue trace plots the modeled spectrum computed with \texttt{proFFiT} 
    using a modulated Voigt profile (see text).
    The red sticks indicates the position and relative intensity of the eighteen 
    hyperfine components.
    Note that the $F_2$ and $F$ quantum numbers are not shown in the plot.}
  \label{fig:413-404}
\end{figure*}

Under our experimental conditions, isolated \ce{NHD} absorption lines have a full width at half 
maximum (FWHM) of 1.5--3.0\,MHz due to the sizable Doppler broadening characteristic of 
this spectral range (e.g., $\Delta\nu^\text{FWHM}_\mrm{G} \sim 1.5$\,MHz at 700\,GHz for \ce{NHD} at 150\,K).
However, due to the complexity of the hyperfine structure, overlap of numerous components is 
common and blends of lines, resolved to different extents, are typically observed.
To facilitate the retrieval of the corresponding central line frequencies, we have modeled 
the recorded absorption profiles using the \texttt{proFFiT} line analysis code 
\citep{Dore-JMS03-proFFit} adopting a FM Voigt profile function and considering 
the full complex representation of the Fourier-transformed dipole correlation function.
When necessary, the background contribution has also been accounted for using a third-order
polynomial expansion.
In most cases, this approach typically allowed for a satisfactory fit of the recorded 
spectral profiles, with some minor discrepancies due to the occasional presence of noise 
features or weak interfering lines.
An example is presented in Figure~\ref{fig:413-404}, which shows the recordings of the  
$N_{K_a,K_c} = 4_{1,3}\leftarrow 4_{0,4}, J = \nicefrac{9}{2}\leftarrow\nicefrac{9}{2}$ 
fine-structure component of \ce{NHD}:
{\bf eighteen} hyperfine components are blended in six resolvable lines illustrating the 
predominance of the N coupling ($I_\mrm{N} = 1$, three main groups), followed by the H coupling 
($I_\mrm{H} = \nicefrac{1}{2}$, closely separated doublets).
The modeled profile has been obtained by fixing the Doppler FWHM at 1.834\,MHz (computed at
150\,K) and adjusting the line positions and the Lorentzian FWHM (collisional broadening),
$\Delta\nu^\text{FWHM}_\mrm{G} = 1.42(15)$\,MHz, assumed equal for all the components.
The described procedure allowed us to obtain line positions with an associate error 
in the 30--100\,kHz interval, depending on the S/N of the spectrum and on 
the overall goodness of the line profile fit.

Whenever possible, a single hyperfine transition has been assigned to a given resolved 
feature detected in the rotational spectrum.
Measurements corresponding to tight line blends have been instead assigned to the subset 
of components which sum up to make the dominant intensity contribution.
In these cases the intensity-averaged frequency is compared with the experimental one
in the least-squares fit.
Loose blends of unresolved components, resulting in very broad or distorted line profiles, 
have not been used in the analysis.

\subsection{Far infrared spectrum} \label{sec:FIR}
\indent\indent
The recorded FIR spectrum was calibrated using residual water lines whose accurate 
frequencies have been reported in the literature \citep{Matsu-JMSt95-H2O}; the resulting 
accuracy on line frequency is estimated to be of 0.0002\,\wn. 
Deuterium scrambling in the \ce{ND3 + H2} discharge mixture appears very efficient as 
strong transitions of \ce{NHD} are observed. 
About 10~of the strongest lines are saturated and have been excluded from the analysis.
As a point of comparison, under the same experimental conditions, in pure \ce{ND3} or \ce{NH3}, 
similarly strong signals of \ce{ND2} and \ce{NH2}, respectively, are observed, while in the 
present spectrum only relatively weak signals of \ce{ND2} (10\,\% absolute absorption 
at best) and  no \ce{NH2} are observed.

\begin{figure*}[t]
  \centering
  \includegraphics[width=0.65\textwidth]{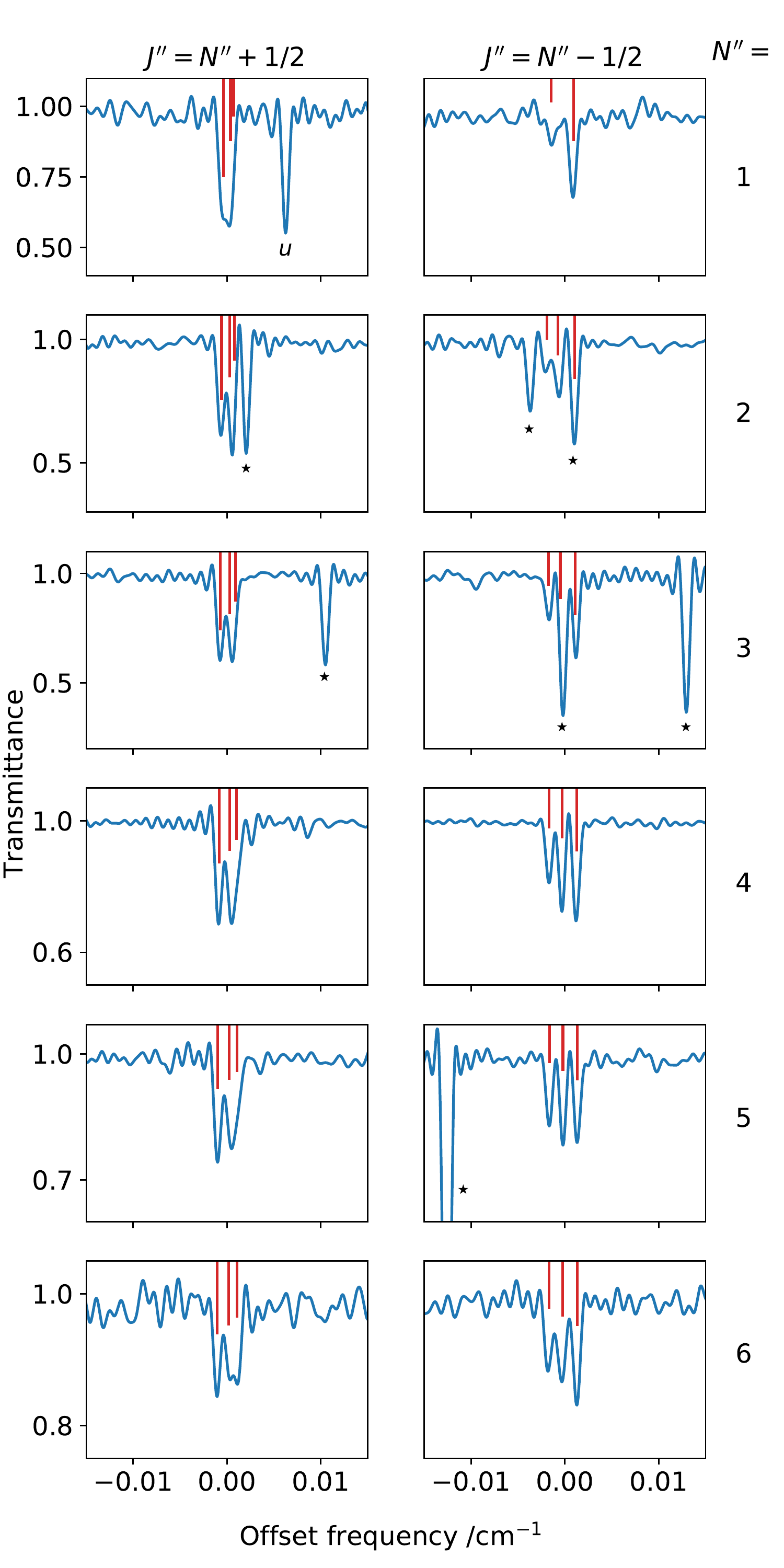}
  \caption{Observation on the FIR spectrum of the hyperfine structure due to the nuclear 
    spin of the N atom in \ce{NHD} for the $b$-type $^rR(1)$ branch with 
    $K_a^{\prime\prime} + K_c^{\prime\prime} - N^{\prime\prime} = 0$. 
    Frequencies are reported as the offset frequencies from the predicted transition 
    when no hyperfine structure is taken into account and span the 68--188\,\wn range 
    (see the supplementary material for the numerical values). 
    Predicted frequency of each component (neglecting the H and D hyperfine splittings) 
    is indicated by red sticks (intensity in arbitrary units). 
    Transitions labelled with ``$\star$'' symbol are still present with the discharge 
    off and most likely arise from \ce{ND3}; the line labelled with a ``$u$'' remains 
    unassigned.}
  \label{fig:NhfsFIR}
\end{figure*}

As previously observed in similar studies on the \ce{NH2} radical and its isotopologues
\citep{Martin-JPCA14-NH2,Margules-AA16-15NH2}, the hyperfine splitting due to the nitrogen 
atom is partially or fully resolved for several $b$-type transitions on the FIR spectrum, 
as illustrated in Fig.~\ref{fig:NhfsFIR}.
Because the natural width of the transitions is smaller than the experimental resolution 
in this spectral region, many lines exhibit a cardinal sine profile at their base 
(arising from the boxcar apodization). 
While this effect can be canceled by applying an apodization to the interferogram 
before Fourier transform, the resulting width broadening prevented from the resolution 
of the nitrogen hyperfine structure on most transitions; consequently such apodization 
was not applied.

In total, 512~transitions have been assigned to 447~experimental frequencies in the range 
54--236\,\wn, 141~$a$-type and 371~$b$-type transitions with 
$N_\text{max}^\prime = 15$ (for $K_a^\prime = 0-2$) and $K^\prime_{a_\text{max}} = 7$ 
(for $N^{\prime}=7$). 
Among these transitions, 107~transitions (80~different frequencies, mostly $b$-type) with 
$K^\prime_a = 0-3$ exhibit partially or fully resolved nitrogen hyperfine structure. 
The uncertainty on line frequency is doubled for lines that appear broadened by 
unresolved hyperfine structure or asymmetric splitting.

\section{Results} \label{sec:res}
\indent\indent
The experimental frequencies were fitted to the Hamiltonian of Eq.~\eqref{eq:Heff} in a 
least-square procedure, in which the weight of each datum was proportional to the inverse
square of its error.
The data set includes previously measured \citep{Morino-JMA97-NH2,Kobay-JCP97-NH2} and 
newly observed mm-, submm-wave, and FIR transitions. 
The data from \citet{Morino-JMA97-NH2} were given uncertainties of $7.5\times 10^{-4}$\,\wn, 
i.e., an order of magnitude smaller than their resolution of 0.0075\,\wn, while the errors 
reported in \citet{Kobay-JCP97-NH2} were retained.
A total of 58~spectroscopic parameters could be determined from the analysis of more than 
1000~distinct frequencies. 
Their values are reported in Table~\ref{tab:rotpar}--\ref{tab:hyppar} and compared with the 
results of former studies.

\begin{deluxetable*}{ll ...}[t]
 \centering
 \tabletypesize{\footnotesize}
 \tablecaption{Rotational and fine-structure spectroscopic parameters determined for \ce{NHD}.
               \label{tab:rotpar}}
 \tablewidth{0pt}
 \tablehead{\mcl{2}{c}{Constants}               & 
            \mcl{1}{c}{Present work}            & 
            \mcl{1}{c}{\citet{Kobay-JCP97-NH2}} & 
            \mcl{1}{c}{\citet{Morino-JMA97-NH2}}}
  \startdata
  $A$            & /MHz & 602874.1082(64)   & 602873.269(18)    & 602879.3(24)    \\
  $B$            & /MHz & 243153.818(15)    & 243150.415(29)    & 243147.6(9)     \\
  $C$            & /MHz & 169872.513(15)    & 169875.877(26)    & 169871.5(9)     \\
  \mcl{5}{l}{Centrifugal distortion} \\
  $\Delta_N$     & /MHz &      9.29361(23)  &      9.2833(9)    &     9.239(9)    \\
  $\Delta_{NK}$  & /MHz &     31.2187(11)   &     30.303(13)    &    31.226(45)   \\
  $\Delta_K$     & /MHz &    250.2294(21)   &    250.825(10)    &   250.90(21)    \\
  $\delta_N$     & /MHz &      3.07038(10)  &      3.04088(12)  &     3.0672(30)  \\
  $\delta_K$     & /MHz &     48.7864(81)   &     46.854(14)    &    47.502(66)   \\
  $\Phi_N$       & /kHz &      1.0098(24)   &                   &     0.785(42)   \\
  $\Phi_{NK}$    & /MHz &      0.04883(14)  &     -0.0901(11)   &     0.04569(60) \\
  $\Phi_{KN}$    & /MHz &     -0.17115(50)  &                   &    -0.1638(21)  \\
  $\Phi_K$       & /MHz &      0.79326(48)  &      0.59213(25)  &     0.8223(87)  \\
  $\phi_N$       & /kHz &      0.4888(12)   &                   &     0.475(18)   \\
  $\phi_{NK}$    & /MHz &      0.023248(49) &                   &     0.02099(42) \\
  $\phi_K$       & /MHz &      0.2588(11)   &                   &     0.1256(19)  \\
  $L_{NNK}$      & /kHz &     -2.373(14)    &                   &                 \\
  $L_{KKN}$      & /kHz &      0.0437(32)   &                   &     0.0615(84)  \\
  $L_K$          & /kHz &     -3.91(29)     &                   &    -3.20(17)    \\
  $l_K$          & /kHz &      0.307(22)    &                   &     0.0210(48)  \\
  $P_K$          & /Hz  &      8.43(29)     &                   &    17.6(16)    \\
  $p_K$          & /Hz  &     -1.40(16)     &                   &                \\
  $O_K$          & /Hz  &     -0.0196(18)   &                   &    -0.0591(57) \\
  \mcl{5}{l}{Fine interaction} \\
  $\epsilon_{aa}$    & /MHz  &                 -7043.123(16)   &  -7043.180(27)  & -7041.0(12)    \\
  $\epsilon_{bb}$    & /MHz  &                 -1149.383(17)   &  -1149.38(9)    & -1150.96(45)   \\
  $\epsilon_{cc}$    & /MHz  &                     7.809(17)   &      7.82(8)    &     8.7(11)    \\
  $(\epsilon_{ab}+\epsilon_{ba})/2$  & /MHz &   1088.64(60)    &   1093.9(11)    &  1056(7)       \\
  $\Delta^S_N$       & /MHz &                      0.12026(78) &      0.1187(20) &     0.140(9)   \\
  $\Delta^S_{KN+NK}$ & /MHz &                      0.2401(87)  &                 &                \\
  $\Delta^S_{NK}$      & /MHz &                      1.77(35)    &      0.234(13)  &                \\
  $\Delta^S_K$       & /MHz &                     16.2441(91)  &     16.245(12)  &    16.16(21)   \\
  $\delta^S_N$       & /MHz &                      0.06300(30) &      0.0640(7)  &     0.075(8)   \\
  $\delta^S_K$       & /MHz &                      0.9553(83)  &      0.96(4)    &     1.20(11)   \\
  $\Phi^S_K$         & /MHz &                     -0.06755(59) &     -0.0634(7)  &    -0.0570(45) \\
  $L^S_K$            & /kHz &                      0.2270(96)  &                 &     0.135(30)  \\[0.5ex]
  \hline
  \mcl{2}{l}{Number of FIR lines}  & \mcl{1}{c}{649}                 \\
  \mcl{2}{l}{Number of MW lines}   & \mcl{1}{c}{353}                 \\
  $\sigma_\text{FIR}$ & /\wn       & \mcl{1}{c}{$4.2\times 10^{-4}$} \\
  $\sigma_\text{MW}$  &  /kHz      & \mcl{1}{c}{51.8}                \\
  \mcl{2}{l}{$\sigma_\text{w}$}  & \mcl{1}{c}{0.963}               \\
  \mcl{2}{l}{$N'_\text{max}$, $K'_{a,\text{max}}$} & \multicolumn{1}{c}{15,9} \\
 \enddata
 \tablenotetext{}{\textsc{Note:} Numbers in parenthesis are one standard deviation in units of the last quoted digit.}
\end{deluxetable*}

\begin{table}
 \centering
 \caption{Hyperfine spectroscopic parameters (MHz) determined for \ce{NHD}.
          \label{tab:hyppar}}
 \begin{tabular}{lc .. }\toprule
  Constant & Atom & \mcl{1}{c}{This work} & \mcl{1}{c}{\citet{Kobay-JCP97-NH2}} \\[0.5ex]
  \hline \\[-1.5ex]
   $a_F$             & (N) &   28.1077(71)  &   28.124(10)  \\[0.5ex]
   $\Delta^K a_F$    & (N) &    0.0         &   -0.0157(12) \\[0.5ex]
   $T_{aa}$          & (N) &  -43.3319(77)  &  -43.329(8)   \\[0.5ex]
   $T_{bb}$          & (N) &  -44.2125(97)  &  -44.218(11)  \\[0.5ex]
   $T_{ab}$          & (N) &    0.550       &    0.550      \\[0.5ex]
   $\chi_{aa}$       & (N) &   -0.288(13)   &   -0.297(14)  \\[0.5ex]
   ($\chi_{bb}-\chi_{cc}$) & (N) &   -6.738(28)   &   -6.725(32)  \\[0.5ex]
   $\chi_{ab}$       & (N) &    1.370       &    1.370      \\[0.5ex]
   $C_{aa}$          & (N) &    0.3486(33)  &    0.3505(37) \\[0.5ex]
   $C_{bb}$          & (N) &    0.0738(29)  &    0.0742(37) \\[0.5ex]
   $\chi_{aa}$          & (D) &    0.167(15)   &    0.143(40)  \\[0.5ex]
   ($\chi_{bb}-\chi_{cc}$) & (D) &    0.100(44)   &               \\[0.5ex]
   $a_F$             & (D) &  -10.230(10)   &  -10.229(13)  \\[0.5ex]
   $T_{aa}$          & (D) &    8.0530(91)   &   8.052(10)  \\[0.5ex]
   $T_{bb}$          & (D) &   -7.251(13)    &  -7.251(14)  \\[0.5ex]
   $T_{ab}$          & (D) &    6.1(14)      &     6.7(14)  \\[0.5ex]
   $C_{aa}$          & (D) &    0.0482(34)   &   0.0491(36) \\[0.5ex]
   $a_F$             & (H) &  -67.128(11)    & -67.116(13)  \\[0.5ex]
   $T_{aa}$          & (H) &   -23.305(31)   &  -23.254(24) \\[0.5ex]
   $T_{bb}$          & (H) &    28.460(23)   &   28.434(22) \\[0.5ex]
   $T_{ab}$          & (H) &    -47.9(19)    &  -48.7(20)   \\[0.5ex]
   $\Delta^K T_{aa}$ & (H) &     -0.122(16)  &  -0.0621(51) \\[0.5ex]
   $\Delta^J T_{aa}$ & (H) &      0.0192(69) &              \\[0.5ex]
   $C_{aa}$          & (H) &      0.0844(75) &   0.1152(64) \\[0.5ex]
   $C_{bb}$          & (H) &      0.0864(50) &   0.1042(60) \\[0.5ex]
   $C_{cc}$          & (H) &     -0.0157(48) &              \\[0.5ex]
  \hline\hline \\[-1ex]
 \end{tabular} \\
 \textsc{Note:} Numbers in parenthesis are one standard deviation in units of the last 
 quoted digit. 
 Parameters without reported deviation were kept fixed in the fit.
\end{table}
  
All the observed transition frequencies could be fitted within their experimental 
uncertainty: the weighted deviation of the fit is close to one ($\sigma = 0.96$) and the 
rms error is 52\,kHz for transitions in the mm-/submm-wave region and 
$4.1\times 10^{-4}$\,\wn for those in the FIR ($1.9\times 10^{-4}$\,\wn, for the transitions 
recorded in this work). 
While the studies of \citet{Morino-JMA97-NH2} and \citet{Kobay-JCP97-NH2}, published almost 
simultaneously, provided two different sets of constants relying on different data sets, our
global analysis comprises all those transitions, in addition to a larger set 
of submm-wave and FIR data.
To this respect, a more comprehensive set of spectroscopic constants is produced in this work.
Overall, the rotational, centrifugal distortion, and fine-interaction constants of 
Table~\ref{tab:rotpar} are now refined, and their values agree well with those 
reported previously by the cited authors.
Still, some sporadic discrepancies are present, e.g., the $\Phi_{NK}$ constant determined 
by \citet{Kobay-JCP97-NH2} is two times larger than ours and has opposite sign.
The hyperfine-interaction parameters are almost identical to those of \citet{Kobay-JCP97-NH2}, 
because the magnitude of the hyperfine coupling effect decreases with the increase of the
rotational quantum numbers involved in the transitions. 
We have determined 23~hyperfine constants, three more than in \citet{Kobay-JCP97-NH2}, namely:
$\chi_{bb}-\chi_{cc}$ for the nucleus D, $\Delta^JT_{aa}$, and $C_{cc}$ for the nucleus H.
The magnitude of these parameters is small and they are determined with a precision of 30\%
at best. 
Still their inclusion in the fit allowed to reduce the $\sigma$ by 10\%, thus they have 
been retained in the final analysis.

While the semi-rigid rotor Hamiltonian used in this work allows to nicely reproduce the 
experimental data at their experimental accuracy, owing to the use of high order centrifugal 
distortion terms, it is possible, however, that the present analysis reaches its limits. 
Indeed, 7 additional transitions involving high $N+K_a$ values have been assigned with 
relative confidence on the FIR spectrum but could not be reproduced in the fit. 
As for \ce{NH2} \citep{Martin-JPCA14-NH2}, a dedicated model taking into account the 
anomalous centrifugal distortion of this very light radical may be better suited to 
reproduce these data. 
Considering the small number of transitions involved, however, no attempt was made in 
this direction in the present study and these transitions are reported in the 
supplementary material.
\clearpage

\section{Conclusions} \label{sec:conc}
\indent\indent
The rotational spectrum of \ce{NHD} has thoroughly been re-investigated from the submm 
region to the FIR.
A frequency-modulation spectrometer and a synchrotron-based FT spectrometer were used to 
perform the measurements.
Numerous $a$-type and $b-$type transitions have been recorded with high accuracy, i.e., 
30--100\,kHz for lines up to 1.2\,THz and 2$\times$10$^{-4}$\,\wn in the 50--240\,\wn 
frequency range.
Also, the astronomically important $N_{Ka,Kc} = 1_{11}-0_{00}$ fundamental $b$-type
transition has been observed in the laboratory for the first time and its fine
and hyperfine components have been accurately measured around 770\,GHz
($J = \nicefrac{3}{2}-\nicefrac{1}{2}$) and 776\,GHz ($J = \nicefrac{1}{2}-\nicefrac{1}{2}$).

The newly obtained set of spectroscopic parameters (Tables~\ref{tab:rotpar} and~\ref{tab:hyppar})
allow to generate a very reliable set of rest-frequencies for \ce{NHD} in the submm and FIR 
spectral domains.
Considering all rotational transitions originating from levels at energy $E/h$ lower than 
300\,K, their corresponding rest-frequencies have a $1\sigma$ uncertainty of less than 
10\,m\,s$^{-1}$ (in radial equivalent velocity) up to 3\,THz.
Two different CDMS-like catalogues based on the present spectral calculations,
with hyperfine structure fully resolved or $^{14}$N coupling only, are provided 
as supplementary material.
This comprehensive laboratory study opens new perspectives for the first identification 
of \ce{NHD} in space, which would enable to make a step towards a more complete 
understanding of ammonia formation in the ISM.
The highly precise rest-frequencies provided here enable to carry out focused searches in the
observation data of Herschel Legacy Archive (\texttt{http://archives.esac.esa.int/hsa/whsa/})
and also provides a basis for new observations in the presently accessible spectral windows
employing facilities like ALMA or the upcoming 4GREAT instrument on board SOFIA.

\section*{Acknowledgments}
\indent\indent
This study was supported by Bologna University (RFO funds) and by MIUR (Project PRIN 2015:
STARS in the CAOS, Grant Number 2015F59J3R).
The work at SOLEIL has been performed under the proposal 20110017 and was supported by 
the Programme National ``Physique et Chimie du Milieu Interstellaire'' (PCMI) of 
CNRS/INSU with INC/INP co-funded by CEA and CNES.

\section*{Supplementary material}
\indent\indent
The catalogue \texttt{nhd\_full\_hfs.cat} lists all the hyperfine components originated by
$^{14}$N, D, and H nuclear couplings and extends up to 2\,THz (150\,\textmu m).
The catalogue \texttt{nhd\_n\_hfs.cat} contains a computation of the hyperfine spectrum
due to the $^{14}$N coupling only and extends into the FIR domain (8\,THz $\sim$ 38\,\textmu m).
The file \texttt{center\_frequencies.dat} reports the reference frequencies used in
Figure~\ref{fig:NhfsFIR}.
The file \texttt{discarded\_transitions.dat} contains those transitions which could
not be reproduced in the fit and were excluded from the analysis.
The machine-readable table \texttt{measured\_transitions.mrt} lists all measured
transition frequencies and fit residuals.

\clearpage


\end{document}